\documentclass[aip, amsmath, amssymb, reprint]{revtex4-1}
\usepackage{graphicx} 
\usepackage{dcolumn} 
\usepackage{bm} 
\usepackage[utf8]{inputenc}
\usepackage[T1]{fontenc}
\usepackage{mathptmx}
\usepackage{etoolbox}
\usepackage{soul}
\usepackage{xcolor}
\usepackage[utf8]{inputenc}
\DeclareUnicodeCharacter{2009}{\,} 
\usepackage{mathtools}
\usepackage{tikz}
\usepackage{tabularx} 
\usepackage{array} 
\usetikzlibrary{tikzmark}

\DeclareMathOperator{\csch}{csch}
\makeatletter
\def\@email#1#2{%
 \endgroup
 \patchcmd{\titleblock@produce}
  {\frontmatter@RRAPformat}
  {\frontmatter@RRAPformat{\produce@RRAP{*#1\href{mailto:#2}{#2}}}\frontmatter@RRAPformat}
  {}{}
}%
\makeatother

\usepackage[most]{tcolorbox}
\newtcolorbox{highlighted}{colback=yellow,coltext=black,breakable}

\begin{document}


\title{Connecting physics to systems with modular spin-circuits}


\author{Kemal Selcuk}
\affiliation{\mbox{Department of Electrical and Computer Engineering, University of California, Santa Barbara, CA, 93106, USA}} 

\author{Saleh Bunaiyan}
\affiliation{\mbox{Department of Electrical and Computer Engineering, University of California, Santa Barbara, CA, 93106, USA}}
\affiliation{\mbox{Electrical Engineering Department, King Fahd University of Petroleum $\&$ Minerals (KFUPM), Dhahran 31261, Saudi Arabia}}

\author{Nihal Sanjay Singh}
\affiliation{\mbox{Department of Electrical and Computer Engineering, University of California, Santa Barbara, CA, 93106, USA}}

\author{Shehrin Sayed}
\affiliation{\mbox{TDK Headway Technologies, Inc.,  Milpitas, CA, 95035, USA}}

\author{Samiran Ganguly}
\affiliation{\mbox{Department of Electrical and Computer Engineering, Virginia Commonwealth University, Richmond, VA 23284, USA}}

\author{Giovanni Finocchio}
\affiliation{\mbox{Department of Mathematical and Computer Sciences, Physical Sciences and Earth Sciences, University of Messina, Messina, Italy}} 

\author{Supriyo Datta}
\affiliation{\mbox{Elmore Family School of Electrical and Computer Engineering, Purdue University, West Lafayette, Indiana, 47907, USA}}

\author{Kerem Y. Camsari}
\affiliation{\mbox{Department of Electrical and Computer Engineering, University of California, Santa Barbara, CA, 93106, USA}}

\begin{abstract}
An emerging paradigm in modern electronics is that of CMOS + $\sf X$ requiring the integration of standard CMOS technology with novel materials and technologies denoted by $\sf X$. In this context, a crucial challenge is to develop accurate circuit models for $\sf X$ that are compatible with standard models for CMOS-based circuits and systems. In this perspective, we present physics-based, experimentally benchmarked modular circuit models that can be used to evaluate a class of CMOS + $\sf X$ systems, where $\sf X$ denotes magnetic and spintronic materials and phenomena. This class of materials is particularly challenging because they go beyond conventional charge-based phenomena and involve the spin degree of freedom which involves non-trivial quantum effects. Starting from density matrices $-$ the central quantity in quantum transport $-$ using well-defined approximations, it is possible to obtain spin-circuits that generalize ordinary circuit theory to 4-component currents and voltages (1 for charge and 3 for spin). With step-by-step examples that progressively become more complex, we illustrate how the spin-circuit approach can be used to start from the physics of magnetism and spintronics to enable accurate system-level evaluations. We believe the core approach can be extended to include other quantum degrees of freedom like valley and pseudospins starting from corresponding density matrices.

\end{abstract}

\maketitle

\section{Introduction}
\label{sec:intro}

The rise of Artificial Intelligence with its skyrocketing computing needs coincided with the stagnation of Moore's Law. This clash has been driving the development of domain-specific hardware \cite{dally2020domain} and architectures with a rich variety of heterogeneous systems for computing, memory and sensing applications. In this new era, rapid and accurate tools for evaluating the potential of emerging materials, physical phenomena, and device concepts has become a crucial need. Such tools will have an impact not only in moving forward well-established computational schemes but also in opening new directions in unconventional computing paradigms \cite{finocchio2023roadmap}.

In this perspective, we describe a physics-based circuit approach that covers a wide range of phenomena in spintronics and magnetism using a generalized circuit theory. We show how circuit ``modules'' derived out of microscopic theory and phenomenological models can accurately model spin transport while accounting for magnetization dynamics.

Combining phenomenology and microscopic theory, the spin-circuit approach for spintronics has been used to model non-local spin-valves, channels with high-spin orbit coupling such as semiconductor channels with Rashba interactions, heavy metals and topological insulators, transport in ferromagnetic insulators, ferromagnet-normal metal interfaces, spin-pumping phenomena, magnetic tunnel junctions, voltage controlled magnetic anisotropy, finite temperature magnetization dynamics and others. A web page with open-source models along with open-source SPICE codes catalogue these results \cite{nanohub:spintronics}. 

The key strength of the approach is not just about modeling phenomena, but more about its ability to \textit{combine} the modules to design \textit{new} circuits and structures. For example, given interface, bulk magnet, magnetization dynamics, and spin-orbit channel modules, complicated new devices can be constructed and studied (see, for example, Ref's.~\cite{sayed2019rectification,sayed2021spin}). Real-time simulation of nanomagnet dynamics coupled with transport modules allows accurate transient simulations from which device characteristics can be obtained. Powerful tools and analysis options of mature circuit simulators greatly ease a wide range of measurements for AC, DC, transient and noise analysis.  Our transport conductances are based on low-frequency (DC) analysis, but they can be dynamically controlled by changing magnetization vectors. We assume that these changes occur instantaneously, allowing the transport to be described using lumped circuit models.

Another distinguishing aspect of spin-circuits compared to powerful alternatives to model spintronic phenomena \cite{,vansteenkiste2014design,mojsiejuk2023cmtj} is how new devices and phenomena can be seamlessly integrated with state-of-the-art complementary metal oxide semiconductor (CMOS) transistor models. This combination allows fast, accurate and  informative evaluation of CMOS + $\sf X$ platforms (where $\sf X$ can be any emerging CMOS-compatible technology such as spintronics, ferroelectrics, photonics, etc.) using efficient circuit simulators (e.g.,  SPICE and its variants). 

The spin-circuit approach evolved out of a 2-component model involving collinear spins \cite{valet1993theory}  which is relatively intuitive. It is as if there are two species of electrons, up and down. The charge current is the sum of up and down currents, while the spin current is given by their difference.

Less intuitive is the 4-component model with noncollinear components, 1 for charge and 3 for spin  (see, for example, \cite{BRATAAStransport, Bauer2003} and references therein). The 4-component model is not based on four species of electrons. Rather, it is based on two components with complex amplitudes $\{u \ v\}$ that embody subtle quantum physics. For example, $\{1 \ 0\}$ represents $+z \ spin$, $\{0 \ 1\}$ represents $-z \ spin$, while a superposition of the two $\{1 \ 1\}$ represents $+x\ spin$. This can lead to quite non-intuitive results, like a flux of $+x \ spins$ getting converted into $+z \ spins$ by a shunt path that pulls out $-z \ spins$\cite{camsari2015modular}. 

\begin{figure*}[t!]
    \centering
    \includegraphics[width=1\textwidth]{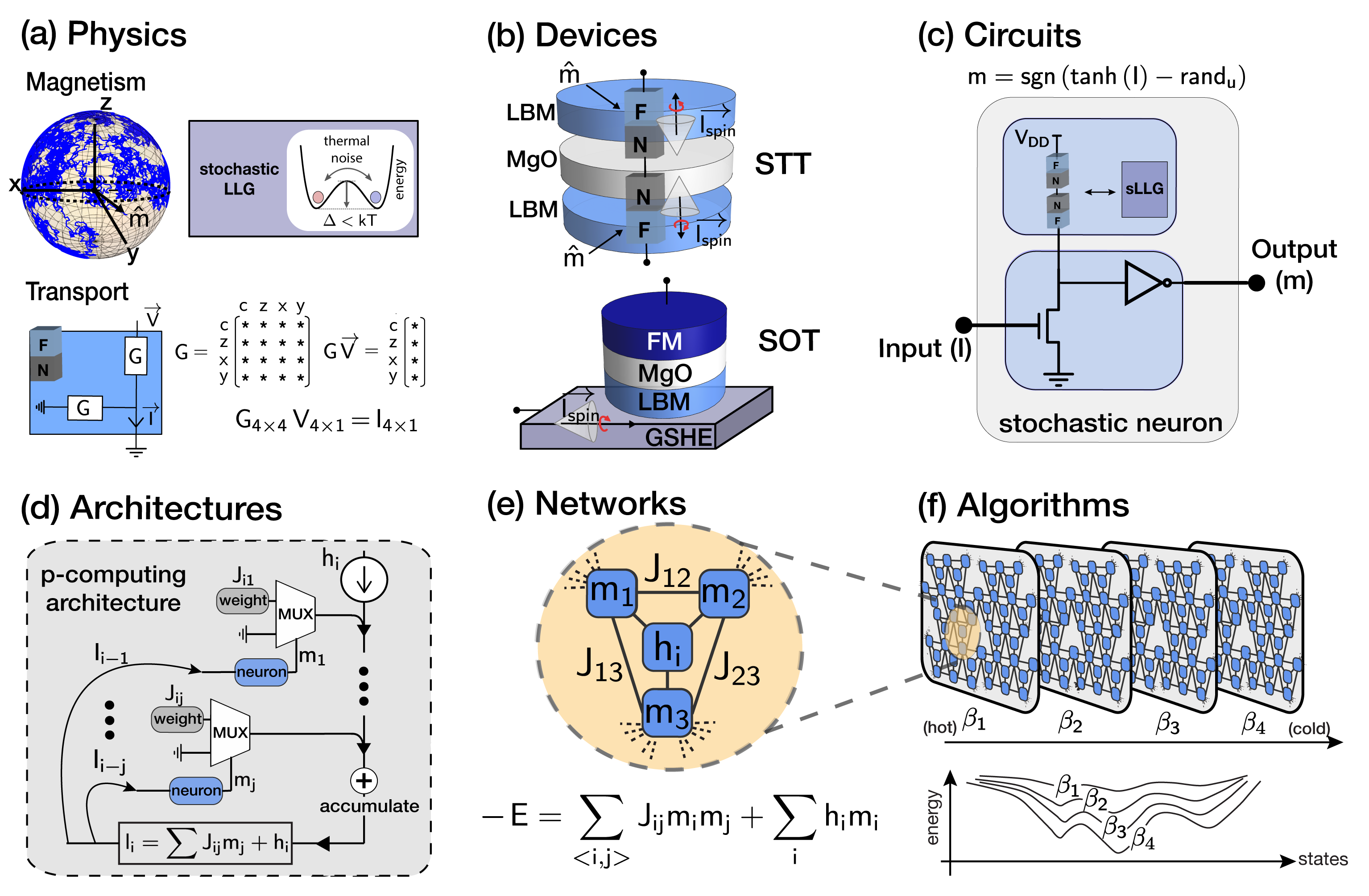}
    \caption{\textbf{Physics to systems perspective with modular spin-circuits}  (a) Physics: Spin-circuits solve transport and magnetization dynamics self-consistently. (b) Devices:  example stochastic MTJs (with spin-orbit and spin-transfer torque) using low-energy barrier magnets. (c) Circuits: stochastic neurons (p-bits) built out of stochastic MTJs (d) Architectures: Probabilistic architectures with interacting stochastic neurons (e) Networks: networks of p-bits mapped to computationally hard optimization problems (f) Algorithms: powerful algorithms that use replicas of probabilistic networks to help solve these optimization problems.
    }
    \label{fi:fig1}\vspace{-15pt}
\end{figure*}

Even such non-intuitive effects are accurately captured by the 4-component model whose components represent measurable quantities given by bilinear products of $u$ and $v$. For the 2-component wavefunction $\psi=\{u \ v\}^T$ with complex components:   
\begin{equation}
\psi \psi^\dagger = \rho= \begin{bmatrix}
\begin{array}{cc}
 u u^* & u v^* \\
 v u^* & v v^* \\
\end{array}
\end{bmatrix}
\end{equation}
where $\rho$ is the density matrix at a given point in the real space representation. The charge and spin components are then given by $\mathrm{tr}(\rho \sigma_i)$ where $\sigma_i$ are Pauli spin matrices for $x$,$y$, $z$ and the identity matrix for charge. Then, the charge component is given by $uu^*+vv^*$ while the three components of spin are given by $uu^*-vv^*$ ($z$-spin), $ 2 \ Re(uv^*)$ ($y$-spin) and $-2 \ Im (uv^*)$ ($x$-spin). 

The 4-component spin-circuit equations have later been converted into  convenient and intuitive 4-component circuits where currents and voltages carry 3-spin and 1-charge components that are related by $4\times 4$ conductances matrices. Many examples of spin-circuits to model existing and evaluate new device concepts have been performed over the years, by the authors and others\cite{srinivasan2011all,srinivasan2013modeling,manipatruni2014vector, manipatruni2012modeling,camsari2015modularA, camsari2016modular,li2021experiments,camsari2014physics, hong2016spin, sayed2016multi,sayed2017proposal, sayed2018transmission,hu2018ac,zand2017energy,roy2017spin,iyengar2016retention,camsari2018equivalent,roy2021spin,hung2019direct,lim2020programmable,li2022physics,thirumala2020valley,cho2021utilizing,roy2023spin}.

We first give a brief introduction to the spin-circuit approach discussing the basics of the transport and magnetism modules and how they interact. To illustrate how extensible and modular the approach is, we present several original examples of the approach by constructing  new spin-circuits. Some of our examples are chosen in the context of a new and emerging computational paradigm with probabilistic bits, covering physics, devices, circuits, architectures, networks, reaching all the way up to the algorithms that run on this stack (FIG.~\ref{fi:fig1}). The ideas related to probabilistic computing came long after the spin-circuit approach but as we will show, spin-circuits have been instrumental in helping uncover new physics and new potential applications due to their modularity enabling a ``plug and play'' approach.


\section{Spin transport with 4-component circuits}
\label{sec:spincircuit}

The two main ingredients in the spin-circuit approach are transport and magnetism modules that need to be solved self-consistently. Transport timescales are typically much faster than magnetization dynamics and this allows a lumped circuit description of transport modules that are solved for each new magnetization configuration in the circuit. We first start by describing spin-transport modules.

Transport modules are naturally represented as circuits but they need to be generalized to include spin transport. If a conductance (or resistance) based formulation for circuit theory is desired, the  principled approach is to start from a quantum transport formulation to obtain relate terminal currents to terminal voltages in terms of conductance matrices. These matrices are of dimension 4$\times$ 4 in the case of spin transport relating 4-component current and voltage vectors, one component for charge and three components for spin directions (we show a concrete example in Section~\ref{sec:rashba}). 

The key point however is that a fully phase-coherent description of conductors is often unnecessary, since spin conductors generally conserve spin information captured in the 2$\times$2 Hermitian part of the density matrix at a real space point, but longer spatial correlations are often irrelevant and they need to be taken out by computationally expensive dephasing mechanisms. As we will show, the spin-circuit approach we discuss can combine \textit{diffusive} spin conductances with \textit{coherent} spin conductances where the coherent part can often be restricted to a small ``active'' region of interest. This effective combination ensures that quantum transport is accounted for only when it is needed. We stress that our examples in this paper are exclusively on spin-transport, but extensions to valley or more complicated degrees of freedom should be possible using similar approaches. 

As we discuss next in Section~\ref{sec:twoport} , the non-conservative nature of spin-currents necessitates care in a circuit description of spin conductances.  These non-conservative currents are naturally handled by shunt conductances that are  connected to grounds. The resulting circuits  fully satisfy Kirchhoff's laws and can be handled by powerful circuit simulators \cite{srinivasan2016magnetic,manipatruni2012modeling,camsari2015modular}.  A microscopic formulation of 4-component formulation of spin-currents was first explored in Ref.'s \cite{ 
BRATAAStransport,Bauer2003}, focusing on metallic and ferromagnetic channels. In our view, however, the spin-circuit formalism is much broader. Even though starting from microscopic theory may not always be necessary or possible, phenomenological 4-component spin-circuit models can still be obtained. Examples of these include spin-circuits for channels with spin-momentum locking\cite{sayed2021unified} such as heavy metals with giant spin Hall effect \cite{hong2016spin}, topological insulators \cite{hong2016spin,sayed2018transmission}, magnonic transport magnetic insulators \cite{sayed2016spin} and others.

\section{Two-port formulation of spin conductances}
\label{sec:twoport}

\begin{figure}[t!]
    \centering
    \includegraphics[width=0.99\linewidth]{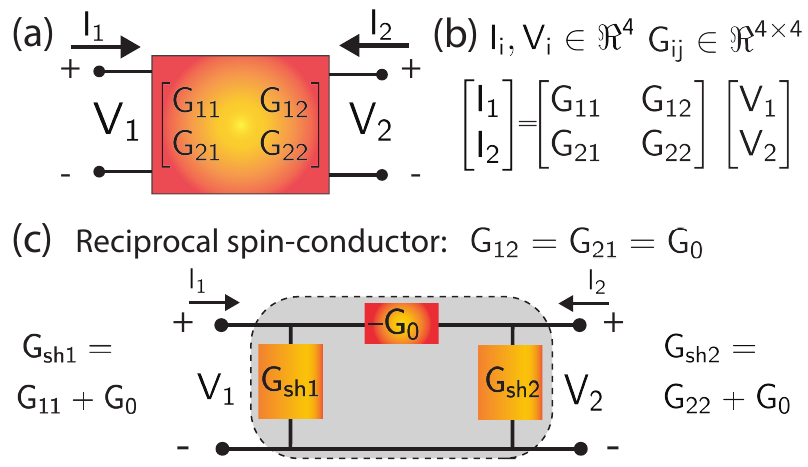}
    \caption{\textbf{2-port formulation of spin conductances} (a) Any 2-terminal spin conductor can be formulated in terms 2-port conductance matrices between its terminals. (b) The currents and voltages are related to each other by 4$\times$4 conductances $ G_{ij}$ and  currents and voltages are 4-component vectors. (c) Unlike charge currents, spin conductors may exhibit non-conservation of currents ($ I_1 + I_2 \neq 0$) and non-reciprocity ($G_{12}\neq G_{21}$). Here, we show an example of a reciprocal spin conductor ($G_{12}=G_{21}=G_0$). Even with non-the conservative nature of spin-currents, it is possible to obtain a circuit description by introducing shunt conductances from the terminal to the ground to account for losses through spin-relaxation or coherent rotation mechanisms.}
    \label{fi:figstar}
\end{figure}

The 4-component conductance formulation is rooted in a 2-port description of transport. For a 2-terminal conductor, the two port formulation relates currents to voltages. In ordinary charge conductors, the 2-port formulation simplifies due to Kirchhoff's current law, which enforces current conservation: $I_1 + I_2 = 0$. This results in constraints like $G_{11} = -G_{21}$ and $G_{22} = -G_{12}$, and if reciprocity holds ($G_{12} = G_{21}=G_0$), only one independent parameter, $G_0$, is needed to fully describe the 2-port conductance matrix. This is why ordinary circuit theory typically does not use a 2-port formulation. For general conductances, neither reciprocity nor current conservation is guaranteed. In spin circuits, unlike charge currents, spin currents are not strictly conserved due to various relaxation processes, such as spin-flip scattering and spin dephasing, or due to coherent rotations from spin-orbit coupling and external magnetic fields. These mechanisms unbalance the spin currents entering and exiting ports, manifesting as $G_{11}\neq -G_{21}$ in the 2-port formulation. Nonetheless, it is \textit{still} possible to represent the 2-port description in terms of a standard circuit with \textit{shunt} conductances (FIG.~\ref{fi:figstar}). When the spin conductance is reciprocal ($G_{12} = G_{21} = G_0$), the system can be represented with shunt conductances $G_{sh1} = G_{11} + G_0$ and $G_{sh2} = G_{22} + G_0$ at each terminal. These shunt conductances capture the losses from spin relaxation or coherent rotations, analogous to how shunt elements handle signal losses and dissipation in microwave circuits \cite{pozar2021microwave}. This reciprocal assumption simplifies the circuit representation, as it allows symmetric handling of currents at both ports. For non-reciprocal spin conductors, additional elements such as dependent sources may be required to capture the asymmetric nature of spin current flow\cite{camsari2019non}. We will examine an example of a non-reciprocal conductance in channels with spin-momentum locking in Section~\ref{sec:rashba}. Modern circuit simulators like HSPICE can also take in the constitutive 2-port relations directly to describe conductances, so both of these representations may be useful.

The 2-port formalism is entirely general and agnostic to where the conductances $G_{ij}$ come from. The examples we consider in this paper cover widely different regimes from coherent quantum  to semi-classical diffusive transport. The conductances can originate from microscopic, phenomenological theory or experiments. 

\begin{figure*}[t!]
    \centering
    \includegraphics[width=1\textwidth]{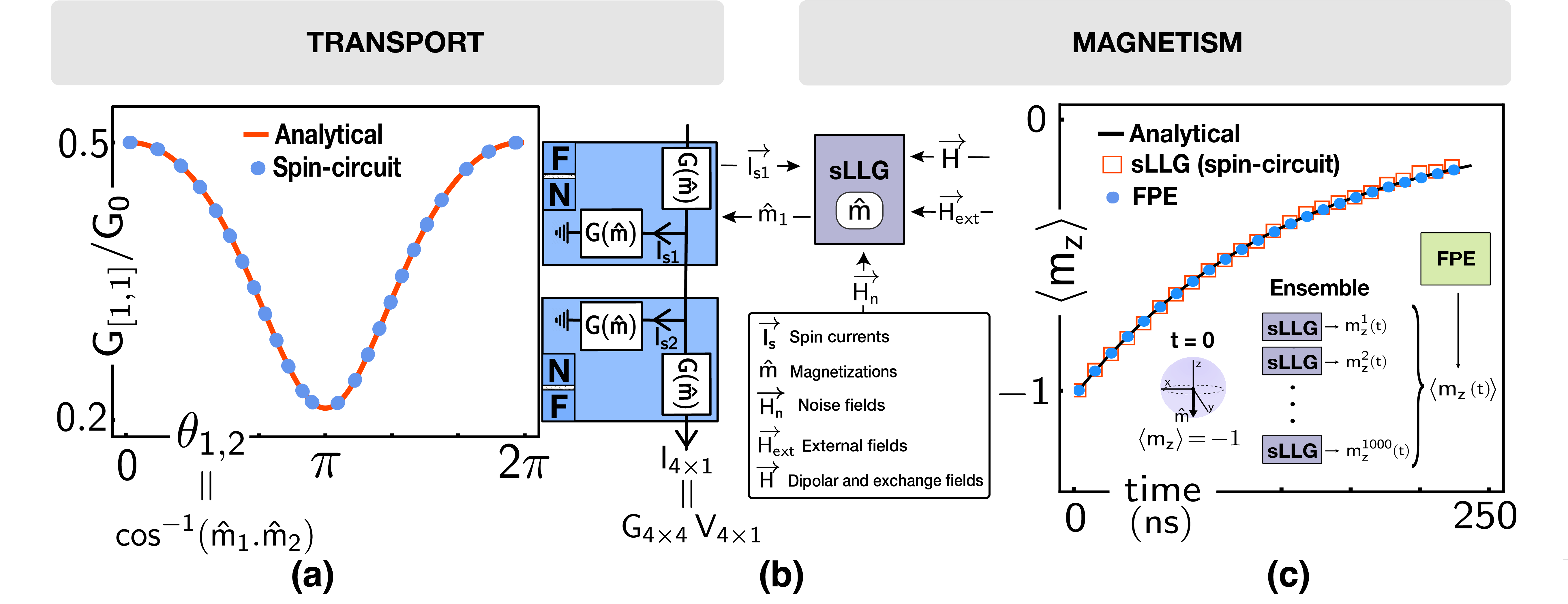}
    \caption{\textbf{Transport and magnetism} (a) An example spin-valve built out of two interfaces is shown. Numerical results obtained from spin-circuits are compared with theory \cite{BRATAAStransport} where the charge conductance shows magnetoresistance as a function of the relative angle between the ferromagnets. (b) Spin-circuit model illustrating the interaction between the magnetization dynamics (modeled by sLLG) and transport modules. The transport model receives two magnetization vectors from the stochastic LLG  and produces 4-component spin currents carrying charge and spin information. sLLG receives spin currents and magnetic fields and produces a magnetization vector. (c) sLLG results are benchmarked with the Fokker Planck equation (FPE). 1000 low-barrier nanomagnets (with a very small perpendicular magnetic anisotropy) are prepared in the $-$1 direction and left to relax. The average magnetization $\left\langle m_z \right\rangle$ is measured over time and compared to  FPE and the analytical solution (see text). 
    }
    \label{fi:fig2}
\end{figure*}
\section{Ferromagnet-Normal Metal Interface}

For many spintronic devices, a key component is the ferromagnet-normal metal interface (F||N) where the spin-transfer-torque effect occurs. A four-component circuit formulation of the F||N interface can be obtained from scattering theory  or the non-equilibrium Green's function formalism \cite{srinivasan2016magnetic,BRATAAStransport,camsari2019non}. The F||N interface consists of a series and a shunt component (FIG.~\ref{fi:fig2}a) that both depend on the orientation of the ferromagnet. When the ferromagnet points in the +z direction, these conductances are given by:
\begin{equation}
\resizebox{\columnwidth}{!}{%
$ \begin{aligned}
G_{se}/G_0 =
\left[ \begin{array}{c|cccc}
  & c & z & x & y \\ \hline
c & 1 & P & 0 & 0 \\
z & P & 1 & 0 & 0 \\
x & 0 & 0&0 & 0 \\
y & 0 & 0 & 0 & 0
\end{array} \right],
G_{sh}/G_0 = 
\left[ \begin{array}{c|cccc}
  & c & z & x & y \\ \hline
c & 0 & 0 & 0 & 0 \\
z & 0 & 0 & 0 & 0 \\
x & 0 & 0 & a & b \\
y & 0 & 0 & -b & a
\end{array} \right]
\end{aligned}
$%
\nonumber
}
\end{equation}
where $G_0$ is the interface conductance, $P$ is the interface polarization, $a, b$ are the real and imaginary coefficients of the ``spin-mixing conductance'', respectively. The form of these conductances is intuitive: the series conductance creates spin-polarized spin currents when subject to a charge potential and shunt conductances are responsible for absorbing transverse spin currents that result in the spin-transfer-torque effect. 

Naturally, in settings where transient behavior needs to be examined,  conductances need to be modified in conjunction with moving ferromagnetic magnetization vectors. This  can be carried out by a standard rotation matrix  that leaves the charge components (cc) unchanged but modifies spin components. Expressing magnetization in spherical coordinates, an arbitrary magnet direction ($\theta$,$\phi$) can be reached via $G_{\{sh,se\}} = [U_R]^T  \left[ G_{\{sh,se\}} \right] [U_R]$, where the rotation matrix $[U_R]$ is given by \cite{camsari2015modular}:
\begin{equation}
\resizebox{\columnwidth}{!}{%
$
\left[ \begin{array}{c|cccc}
& c & z & x & y \\ \hline
c& 1&0&0&0\\
z&0& \cos \theta & \sin \theta \cos \phi & \sin \theta \sin \phi \\
x&0& -\sin \theta \cos \phi & \cos \theta + \sin^2 \phi (1 - \cos \theta) & -\sin \phi \cos \phi (1 - \cos \theta) \\
y&0& -\sin \theta \sin \phi & \sin \phi \cos \phi (1 - \cos \theta) & \cos \theta + \cos^2 \phi (1 - \cos \theta)
\end{array} \right]
$%
}\nonumber 
\end{equation}

In circuit simulators, we first obtain a fully paramaterized rotated conductance that receives instantaneous magnetization directions for transient simulations.  

As a simple example that demonstrates the modularity of such spin-circuits, 
FIG.\ref{fi:fig2}(a) shows a metallic spin-valve where the relative angle between the ferromagnets is changed. In this case (and in many cases involving spin-circuits) the charge conductance (or the c-c component) of the equivalent conductance can be analytically calculated (see Eq.~124 in  \cite{BRATAAStransport} with $a = 2 \Re\left(G_{\uparrow \downarrow}/G_0\right)$, while the imaginary part $b$ is set to 0, which is typical for metallic interfaces). FIG.~\ref{fi:fig2}(a) shows the magnetoresistance effect on the charge conductance where the analytical result is compared to a numerical one obtained from a circuit simulator (HSPICE). 

This example shows the magnetoresistive change in the charge conductance, but the spin-circuit also captures important spin current information that can be readily extracted. Technically, what we illustrate here is a metallic spin-valve. Remarkably, \textit{multiplying} two conductance matrices instead of adding them in series seems to capture the non-trivial magnetic tunnel junction physics \cite{camsari2014physics}. The intuition behind this is the exponential decay of conductance across two tunneling interfaces in series $G_{eq} \propto G_1 G_2$ which seems to generalize to matrix conductances. 

The spin-valve example we show in FIG.~\ref{fi:fig2} may seem elementary however the approach is more general. Recently, Ref.~\cite{kemal2024double} analyzed a complicated magnetic tunnel junction design with two synthetic antiferromagnetic layers (4 ferromagnets) using  the same approach, obtaining results in agreement with experimental features observed in similar systems \cite{sun2023easy}.

\section{Channels with spin-orbit coupling}  
\label{sec:rashba}  

Other than the FM|NM interface, the transport conductances we consider in this paper are generally based on 4-component spin-diffusion equations. As another example of how \textit{coherent} quantum transport can be distilled into spin-circuits, we now examine channels with spin-orbit coupling. Consider the following Hamiltonian with Rashba and Dresselhaus terms for a 2D semiconductor\cite{liu2006persistent}:
\begin{equation}
H = H_0 + \alpha (\sigma_x k_y - \sigma_y k_x) + \beta (\sigma_x k_x + \sigma_y k_y)
\label{eq:hamiltonian}
\end{equation}
Here, $H_0$ represents the kinetic energy term of the electrons in the 2D electron gas (2DEG), typically described as: $H_0 = {\hbar^2 k^2}/{2m^*}$
where $\hbar$ is the reduced Planck’s constant, $k$ is the wavevector, and $m^*$ is the effective mass of the electrons in the 2DEG. The terms $\alpha$ and $\beta$ denote the strength of the Rashba and Dresselhaus spin-orbit coupling, respectively. These terms lead to spin-momentum locking, where the effective magnetic fields seen by the electron depends on its momentum. Given this microscopic Hamiltonian, it is possible to derive 4-component 2-terminal conductances required for the 2-port formulation using the Non-Equilibrium Green's Function (NEGF) formalism \cite{camsari2019non,camsari2022nonequilibrium}: 
\begin{align}
\left[{G}_{mn}\right]^{\alpha \beta} &= \frac{q^2}{h} \, 
\text{tr} \left[ i \left( S_{\beta} S_{\alpha} G^R \Gamma_m 
- S_{\alpha} S_{\beta} G^A \Gamma_m \right) \delta_{mn} \right] \notag \\
&\quad - \text{tr} \left[ S_{\alpha} \Gamma_m G^R S_{\beta} \Gamma_n G^A \right]\label{eq:negf}
\end{align}
\noindent where $\left[{G}_{mn}\right]^{\alpha \beta}$ denotes the conductance matrix element between terminals $m$ and $n$ for $\alpha$ and $\beta$ that go over charge and spin $(z,x,y)$. The prefactor ${q^2}/{h}$ involves the electron charge $q$ and Planck's constant $h$. The trace operation, denoted by $\text{tr}$, is taken over spin indices. Here, $G^R$ and $G^A$ are the retarded and advanced Green's functions, respectively, and $\Gamma_m$ and $\Gamma_n$ are the broadening matrices at terminals $m$ and $n$. The matrices $S_{\alpha}$ and $S_{\beta}$ are spin projection matrices corresponding to the spin components, including charge, $z$, $x$, or $y$ spins. The Kronecker delta, $\delta_{mn}$, ensures that the first term contributes only when $m = n$. Eq.~\ref{eq:negf} can be considered the spin-generalization of well-known Landauer formula, obtained from NEGF. In the Appendix, we show, numerically and analytically, that for a 1D ballistic conductor ($k_y=0$), at a conducting energy, the Hamiltonian of Eq.~\ref{eq:hamiltonian} results in:  $G_{11}=G_{22}= ({2 q^2}/{h}) I_{4\times 4}$ where $G_0$ is ${2 q^2}/{h}$ and $-{G_{12}}/{G_0}$ (in the $c,z,x,y$ basis) is: 
\begin{equation}
\resizebox{\columnwidth}{!}{%
$\begin{aligned}
\left[ 
\begin{array}{cccc}
1 & 0 & 0 & 0 \\
0 & \cos\theta & \cos\gamma\sin\theta & \sin\gamma\sin\theta \\
0 & -\cos\gamma\sin\theta & \cos^2\gamma\cos\theta+\sin^2\gamma & -\sin(2\gamma)\sin^2(\frac{\theta}{2}) \\
0 & -\sin\gamma\sin\theta & -\sin(2\gamma)\sin^2(\frac{\theta}{2}) & \sin^2\gamma\cos\theta+\cos^2\gamma \\
\end{array} 
\right]
\end{aligned}
$%
}
\label{eq:gsoc}
\end{equation}

\noindent where we introduced $\gamma=\tan^{-1}(\beta/\alpha)$ and $\theta = \sqrt{\alpha^2 + \beta^2}{(2 m^*L)}/{\hbar^2}$, for a channel length of $L$. It is easy to check that for $\beta=0$, this conductance expresses  coherent precession around the $y$-axis and for $\alpha=0$, it expresses coherent precession around the $x$-axis. Assuming periodic boundary conditions across the width of the sample, it is also possible to include transverse modes ($k_y \neq 0$) to get an averaged out conductance for 2D channels, but we do not attempt this here\cite{zainuddin2011voltage,camsari2019non}. Alternatively, a direct 2D NEGF calculation with fixed boundary conditions can be used to derive the conductance matrix using Eq.~\ref{eq:negf}. 

Another interesting aspect is the non-reciprocity of spin conductances naturally arising in systems with spin-momentum locking. Applying Eq.~\ref{eq:negf} to get $G_{21}$ results in a conductance matrix where $\theta$ is replaced by $-\theta$, due to the momentum dependent effective magnetic fields induced by spin-orbit terms.  These conductances can then be used in circuit simulators to model coherent active regions with spin-orbit coupling, along with FM|NM interfaces that describe magnetic contacts which can then be combined with self-consistent magnetization modules. All of this makes analyzing practical devices such as the Datta-Das transistor\cite{koo2009control} or persistent spin helix states (when $\alpha=\beta$\cite{liu2006persistent}) much more convenient than a full coherent quantum transport treatment.  

Eq.~\ref{eq:negf} assumes coherent conductance over a length of $L$. Therefore, spin-orbit conductances to describe a conductor of length $(2L)$ cannot be obtained by combining two conductances in ordinary circuits, and a new coherent conductance description over $(2L)$ is needed. Interestingly however, \textit{multiplying} the rotation submatrix of two conductances in series achieves a rotation of $(2\theta)$ about the rotation axis, which is what would be obtained from a coherent description of a channel length of $(2L)$. This is reminiscent of multiplied FM|NM conductances to get the correct MTJ physics rather than metallic spin-valves whose physics can be obtained by inverting the conductance matrices adding them in series to get the equivalent 4$\times$4 resistance matrix. The multiplication trick could allow an effective spin-diffusion theory of coherent 4$\times$ 4-conductances (see an alternative direct attempt to obtain a diffusive quantum theory of Rashba SOC in Ref.~\cite{barletti2010quantum}), in networks representing arbitrary geometries. Unfortunately, however, the multiplication of conductances in series is not amenable to standard circuit theory. 

We presented a specific example of channels with spin-orbit coupling, however the NEGF formulation of Eq.~\ref{eq:negf} is entirely general and can produce spin conductances for other types of systems starting from microscopic Hamiltonians. 

\section{Magnetization dynamics via Landau-Lifshitz-Gilbert equation}

For device analysis, the transport captured by spin-circuits typically needs to be solved self-consistently with magnetization dynamics.  Large ferromagnets in experiments typically contain many domains and to get realistic dynamical behavior, sophisticated ``micromagnetics'' tools need to be used \cite{vansteenkiste2014design}. These tools  solve partial differential equations that are hard to combine with circuit simulators. Our approach is to assume monodomain magnets and use the stochastic Landau-Lifshitz-Gilbert (sLLG) equation to model magnetization dynamics. This approximation gets better as magnets are scaled down to small dimensions but more importantly, it allows magnetism and transport modules to be readily coupled in circuit simulators. Moreover, as we show in Section III, multiple monodomain LLG modules can be combined to describe multi-domain physics of nanomagnets, in principle. The single sLLG model incorporates finite temperature physics, dipolar and exchange coupling and spin-transfer torques. 

The sLLG equation is a non-linear 2-dimensional ordinary differential equation  where the magnetization evolves on the surface of the unit sphere \cite{sun2004spin,sun2000spin,butler2012switching,lopez2002transition, amentnumericalLLG}:
\begin{equation}
\begin{aligned}
(1 + \alpha^2)\frac{d\mathbf{\hat{m}}}{dt} = &-|\gamma| \mathbf{\hat{m}} \times \vec{H} - \alpha|\gamma| \mathbf{\hat{m}} \times (\mathbf{\hat{m}} \times \vec{H}) \\
&+ \frac{\alpha}{qN}(\mathbf{\hat{m}} \times \vec{I}_s) + \frac{1}{qN}(\mathbf{\hat{m}} \times (\vec{I}_s \times \mathbf{\hat{m}}))
\end{aligned}
\label{eq:sLLG}
\end{equation}
where $\alpha$ is the damping coefficient, $q$ is the electron charge, $\gamma$ is the electron gyromagnetic ratio, $\vec{I}_s$ is the received spin current. $N$ is the total number of spins in the free layer, $N = M_s \text{Vol.}/\mu_B$, where $M_s$ is the saturation magnetization, $\mu_B$ being the Bohr magneton. In addition to all the fields (uniaxial, demagnetization, external  magnetic fields, strain-induced anisotropy fields, etc.) that go into the effective field $\vec{H}$, the effect of thermal noise also enters as a fluctuating magnetic field with the following properties: 

\begin{equation}
\mathsf{Var.} (H_n^{x, y, z}) = \frac{2\alpha k_B T}{|\gamma| \mu_0 M_s \text{Vol.} }, \quad  
  \mathbb{E}[H_n^{x,y,z}] = 0 
\label{eq:llgnoise}
\end{equation}
where $T$ is the temperature, $k_B$ is the Boltzmann's constant and $\mu_0$ is free space permeability. The noise is assumed to be independent in all 3-dimensions. \\ 

To solve the sLLG equation using powerful circuit simulators, we express the LLG equation in a form of coupled capacitors:\ $ C{dV}/{dt}  = I $\cite{panagopoulos2013physics,camsari2015modularA} where the voltages map to magnetizations and non-linear current sources map to the different terms in the LLG equation. Note that our approach does not use linearization or make any approximation: through the use of non-linear and state-dependent current sources, the \textit{full} LLG equation is solved in circuit simulators. The numerically challenging transient noise simulations can be handled by reformulating existing noise models that are used for resistor noise in HSPICE \cite{torunbalci2018modular}. 

Solving the stochastic LLG requires care, especially if done in closed-source circuit simulators. The time dependence of noise fields, the choice of convention in integration (Itô vs Stratonovitch), and the way the variance of the noise enters in HSPICE may not be obvious. Our approach to such uncertainties is to rigorously benchmark the sLLG by its corresponding Fokker-Planck equation (FPE) \cite{li2004thermally,xie2016fokker,butler2012switching}. For a magnet with cylindrical symmetry, the time-dependent FPE reads\cite{butler2012switching}: 

\begin{figure*}[t!]
    \centering
    \includegraphics[width=1\textwidth]{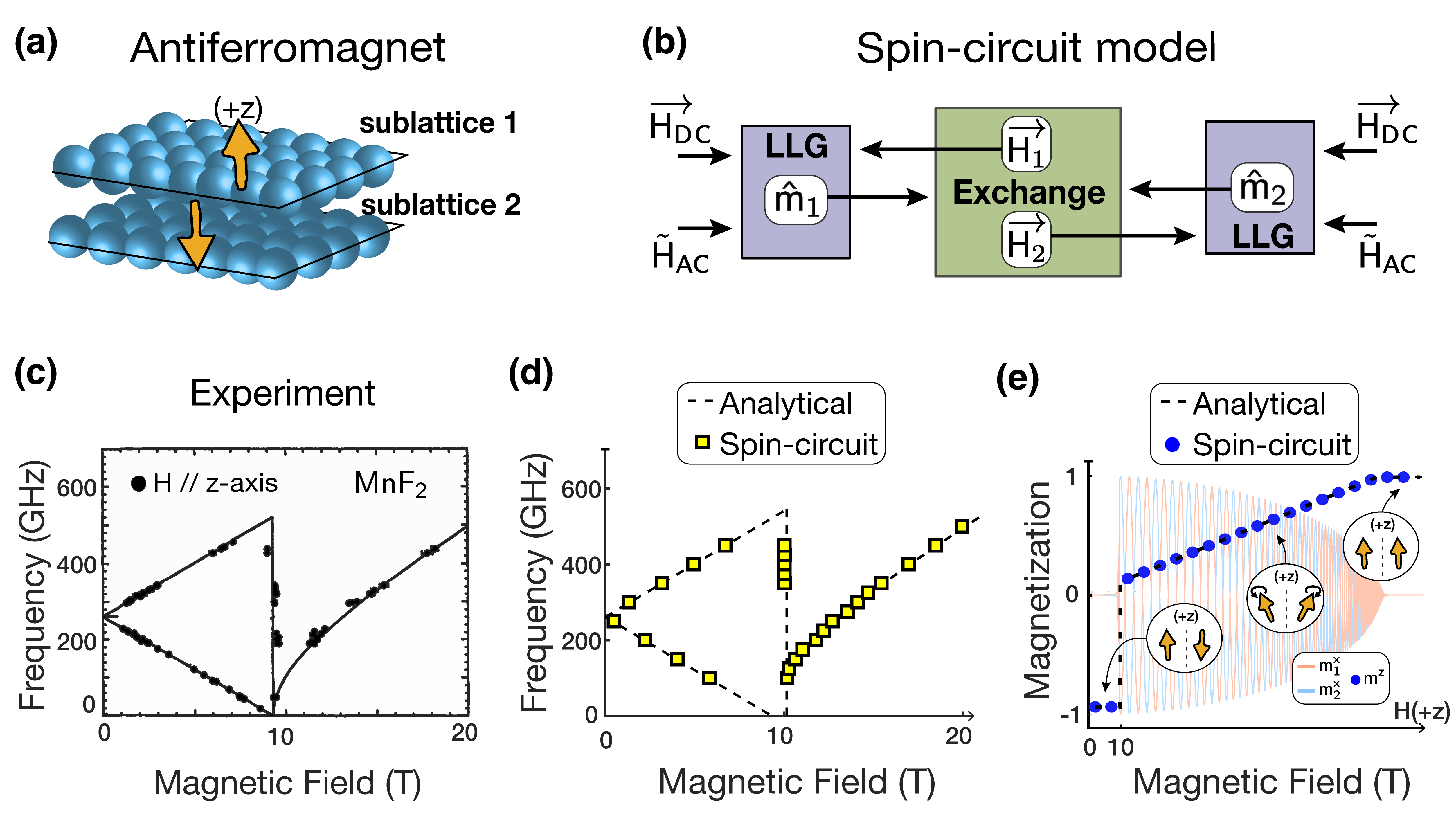}
    \caption{\textbf{Antiferromagnetic resonance (AFMR) with spin-circuits}  (a) Sketch of an antiferromagnet where two sublattices with opposing magnetizations. (b) Spin circuit model of the AFM with two antiferromagnetic layers analyzed by two LLGs coupled with exchange interactions.  (c) Experimental results for AFMR in $\sf{MnF_2}$ \cite{afmrexperiment}. (d) Numerical results obtained from spin-circuits for AFMR, compared to known theory. 
     (e) Easy-axis (z) component of the magnetization vector analysis over an external magnetic field applied in the +z direction. At a critical field, the sublattice spins enter the ``spin-flop'' region where they both develop a small $m_z$ component in the direction of the  magnetic field. In all cases, spin-circuits provide excellent agreement with known theory.  
    }
    \label{fi:fig3}
\end{figure*}
\begin{equation}
\frac{\partial \rho(m_z, t)}{\partial \tau} = \frac{\partial}{\partial m_z} \left[ (i - h - m_z)(1 - m_z^2)\rho + \frac{1 - m_z^2}{2\Delta} \frac{\partial \rho}{\partial m_z} \right]
\label{eq:1DFPE}
\nonumber 
\end{equation}
as long as the external fields $h$ and spin currents $i$ are defined to be in the $\pm z$ direction. $\tau$ is the normalized time, $\tau = (1 + \alpha^2)/( \alpha \gamma H_k) / t$, where  $\alpha$ is the damping coefficient, $H_k$ is the uniaxial anisotropy constant, $\gamma$ is the gyromagnetic ratio, and $t$ is the real time. $\Delta$ represents the energy barrier of the magnet normalized with $k_B T$. 

To benchmark our sLLG solver in HSPICE with FPE, we consider an ensemble of low-barrier nanomagnets all prepared in the $m_z=-$1 direction, which are then left to fluctuate on their own in the absence of any fields and currents. We perform 1000 independent (with identical parameters) sLLG simulations and numerically plot the average $m_z$ component as a function of time. The same quantity can be obtained from a numerical solution of the FPE, which solves for $\rho(\tau,m_z)$. We then integrate $\rho$ to obtain $\left\langle m_z (\tau) \right\rangle = \int \rho(m_z, \tau)\, dm_z$. The FPE sLLG comparison is shown in FIG.~\ref{fi:fig2}c with excellent agreement. Further, both numerical methods can be compared to an analytical expression for the average $m_z$ (following a similar approach in \cite{hassan2019low}: 
\begin{equation}
C(t) = \exp\left(-2\alpha\gamma \frac{k_B T}{M_s Vol.} |t|\right)
\label{eq:PMAautocorr}
\end{equation} 
Eq.~\ref{eq:PMAautocorr} is also shown {as the analytical solution} in FIG.~\ref{fi:fig2} in agreement with FPE and sLLG. These toy examples demonstrate the validation of our numerical solvers by matching the stochastic Landau-Lifshitz-Gilbert (sLLG) simulations with the Fokker-Planck Equation (FPE), which are further compared against analytical predictions.

\begin{figure*}[t!]
    \centering
    \includegraphics[width=1\textwidth]{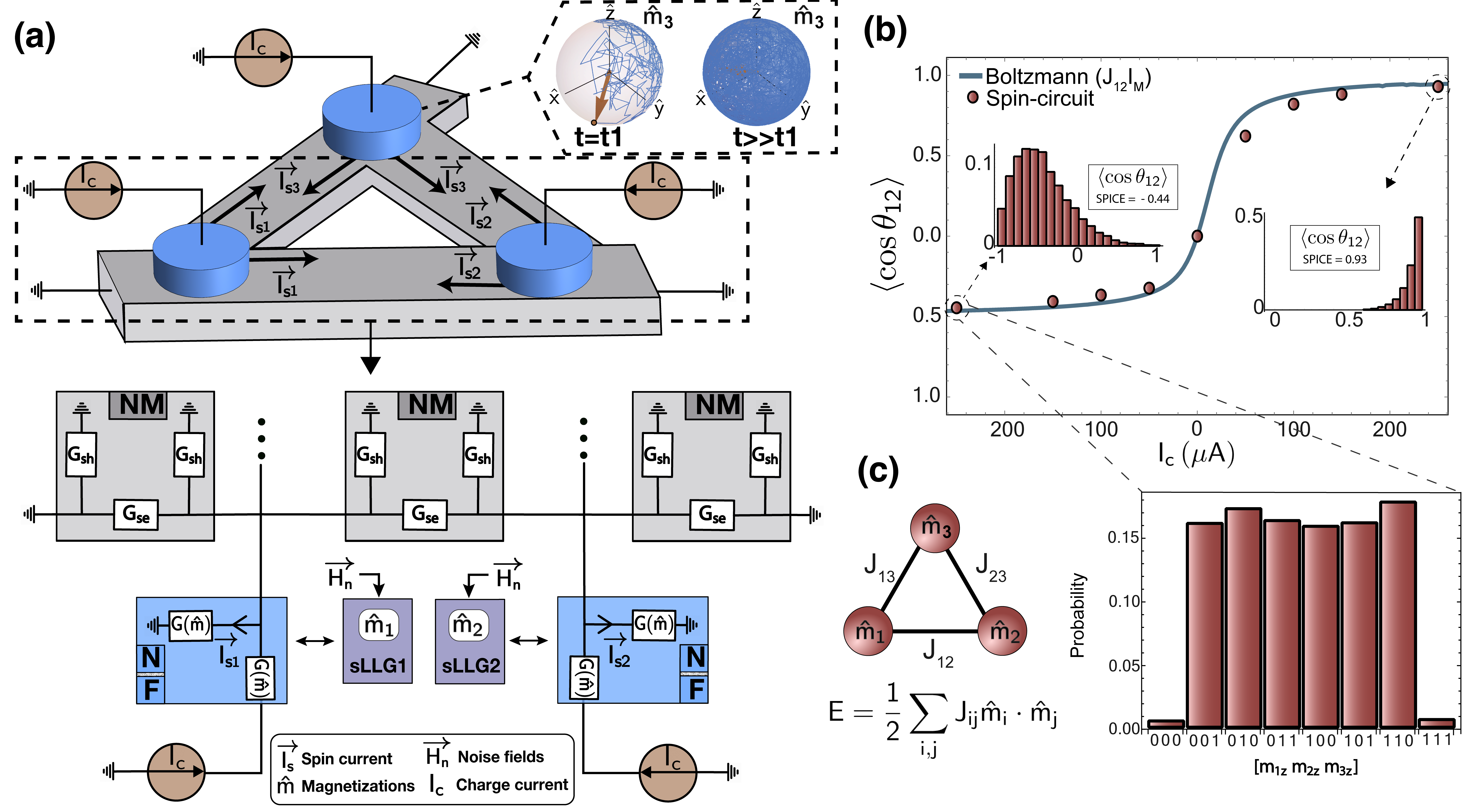}
    \caption{\textbf{Non-local spin valve (NLSV) with low-barrier nanomagnets (LBM)} (a) Physical structure consisting of networks of LBMs. A  charge current is injected from LBMs going to a nearby local ground. Spin currents polarized in the direction of fluctuating LBMs are routed to one another.  Inset shows an example of how magnetization dynamics $\hat m$ evolve over time for an LBM with low perpendicular anisotropy. The bottom panel shows the spin-circuit corresponding to the physical structure. (b) The average of the relative angle between LBM 1 and LBM 2 is measured as a function of injected charge currents, showing ferromagnetic (at positive $I_c$) and antiferromagnetic (at negative $I_c$) coupling. The numerical results are compared with those obtained from the Boltzmann law obtained from the Heisenberg Hamiltonian. This correspondence between the unitless Heisenberg Hamiltonian and spin-circuit requires a mapping factor $I_M$ with units of currents (see text and Ref.~\cite{bunaiyan2023heisenberg}). (c) A histogram of three LBMs at large negative currents where for better illustration the magnetizations $\hat m$ are binarized by thresholding at $\hat m_z =0$. The system shows frustration in the antiferromagnetic configuration.
    }
    \label{fi:fig4}
\end{figure*}

\section{Natural antiferromagnets with spin-circuits}
\label{sec:AFMR}

As mentioned earlier, our approach with magnetization dynamics necessarily assumes the monodomain approximation. Could spin-circuit models be built out of coupled magnetizations? We answer this question in the context of natural antiferromagnets (AFM) by matching the experimental antiferromagnetic resonance behavior observed in MnF$_2$\cite{afmrexperiment}. FIG.~\ref{fi:fig3}a-b shows the two coupled atomic sublattices and the corresponding spin-circuit model. The exchange fields between two magnets can be obtained from an energy model of the form: \cite{victora2005composite}
 \begin{equation}
    E_{ex} = -M_s (\text{Vol}_1 + \text{Vol}_2) ({J}_{ex}) (\hat{m_1} \cdot \hat{m_2})
    \label{eq:Benergyreduced}
\end{equation}
where the effective field that enters the LLG becomes:
\begin{equation}
\left[H_{ex}\right]_i=-\frac{1}{\left(M_s\right)_i V_i} \nabla_{\hat{m}_i} ,\  E_{ex} =  J_{ex} \frac{(\text{Vol}_1 + \text{Vol}_2)}{\text{Vol}_i} m_j 
\label{eq:exchfield}
\end{equation}
In the spin-circuit, the exchange fields are assumed to change ``instantaneously'' for the two coupled LLGs. The tools available to circuit simulators make measuring the AFMR frequency highly convenient. We apply an external DC magnetic field along the z-axis and sweep the frequency of an external AC magnetic field (perpendicular to the z-axis)  and measure the transient response of the $m_z$ components of the constituent spins. The frequency of the AC field at which this response is maximum is recorded as the AFM resonance frequency at that DC field. Interestingly, this is not too different from how the AFMR is measured experimentally. 

By linearizing the coupled LLG equations, two sets of AFMR frequencies can be obtained\cite{kefferAFMR,rezendeAFMR}: 
\begin{eqnarray}
f_\text{res} = \gamma \left(\sqrt{H_{K} \cdot (H_{K} + 2 \cdot J_{\text{ex}})} \pm H_\text{ext} \right)
\label{eq:AFMRreg1down}\\
f^{\sf sf}_\text{res} = \gamma \sqrt{H_{\text{ext}}^2 - (2 \cdot J_{\text{ex}} \cdot H_{K} + H_{K}^2)}  
\label{eq:AFMRreg3}
\end{eqnarray}
where $H_K$ is the uniaxial anistropy of individual sublattices, $H_{ext}$ is the external magnetic field and $\gamma$ is the gyromagnetic ratio for the electron. In FIG.~\ref{fi:fig3}c-d, we observe that around 10 T, the coupled AFM spins enter the interesting ``spin-flop'' region where they each develop a small $m_z$ component and start precessing about this axis (FIG.~\ref{fi:fig3}e).

As FIG.~\ref{fi:fig3} shows, all of this physics is captured by the spin-circuit formalism quantitatively. The availability of AC/DC sources, transient and AC simulation options offer a convenient platform to study magnetization physics in a modular manner. The ability to combine such magnetic models with materials and transistors makes the spin-circuit approach appealing. 

\begin{figure*}[t!]
    \centering
\includegraphics[width=0.95\textwidth]{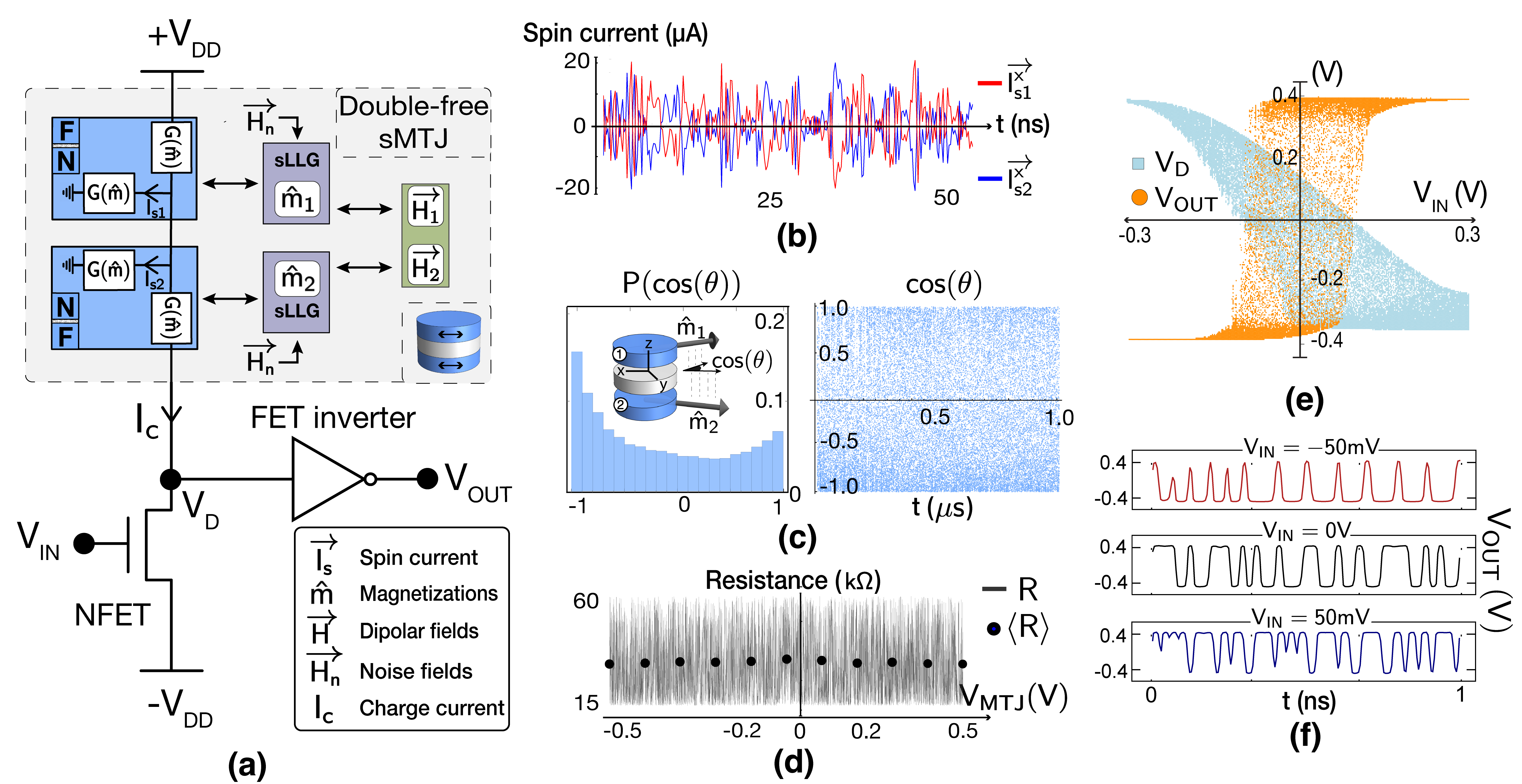}
    \caption{\textbf{Probabilistic bit with double-free-layer stochastic MTJs} (a) Self-consistent magnet and transport model combined with transistors to model a probabilistic bit. (b) Time-dependent spin currents are produced from the transport model that goes into the sLLG modules. We show the x-axis component of spin-currents for magnets, 1 and 2. (c) Histogram and time fluctuations for the $\cos(\theta)$ between $m_z$ components of magnet 1,2 for the double-free-layer sMTJ. Slight anti-parallel tendency is due to the dipolar coupling which is not completely overcome by thermal fields. 
    (d) Resistance of the sMTJ is measured while the voltage is swept from $-$0.5 to 0.5 V over 1 ms. The discrete data points are average resistances over 500 ns showing the roughly bias-independent characteristics of the device. (e) The drain node ($\sf{V_D}$) and the output of the inverter ($\sf{V_{OUT}}$) are measured while the input ($\sf{V_{IN}}$) is swept from $-$0.3 V to 0.3 V over 1 ms. The output of the inverter shows the binary stochastic neuron behavior. (f) Digital output fluctuations over time for the probabilistic bit output at different bias conditions for $\sf{V}_{IN}$.
    }
    \label{fi:fig5} 
\end{figure*}

\section{engineered antiferromagnets with non-local spin valves}
\label{sec:NLSV}
Next, we show an example non-local spin valve (NLSV) setup that couples low-barrier nanomagnets (LBMs) to engineer a ``Heisenberg machine'' using spin-circuit models. The setup we consider here has recently been proposed theoretically \cite{bunaiyan2023heisenberg} and to the best of our knowledge no experiments involving LBMs and NLSVs have yet been performed. Our main point is that modular spin-circuits can be useful to motivate new experiments, provide quantitative insights into new physics and estimate energy and delay metrics before the physical realization of a proposed system. 

In previous sections, we described how we can model magnets by coupling the transport model (F||N) with magnetization dynamics obtained through  the sLLG equation. For channel materials used in NLSVs, we now introduce the Normal Metal (NM) model, describing spin-diffusion in channels without any spin-orbit coupling. The NM model consists of a $\pi$-network with two shunt conductance $G_{sh}$ separated by a series conductance $G_{se}$ as shown in FIG.~\ref{fi:fig4}\cite{srinivasan2013modeling}:

 \begin{equation}
 G_{se}= 
 \left[ \begin{array}{c|cccc}
   & c & z & x & y \\ \hline
 c & G_c & 0 & 0 & 0 \\
 z & 0 & G_s & 0 & 0 \\
 x & 0 & 0&G_s & 0 \\
 y & 0 & 0 & 0 & G_s
 \end{array} \right]
 ,\quad 
 G_{sh}= 
 \left[ \begin{array}{c|cccc}
   & c & z & x & y \\ \hline
 c & 0 & 0 & 0 & 0 \\
 z & 0 & G_s' & 0 & 0 \\
 x & 0 & 0&G_s'&0 \\
 y & 0 & 0 & 0 & G_s'
 \end{array} \right]
 \nonumber 
 \end{equation}
 where $G_c = A_{NM}/(\rho_{NM}L)$, $G_s = A_{NM}/(\rho_{NM}\lambda_s)\csch(L/\lambda_s)$, and $G_s'= A_{NM}/(\rho_{NM}\lambda_s)\tanh(L/2\lambda_s)$. $A_{NM}$ denotes the area, $\rho_{NM}$ is the NM resistivity, $L$ is the length, and $\lambda_s$ is the spin-diffusion length. These matrices are obtained from microscopic spin-diffusion equations, accounting for the non-conservative nature of spin currents through the shunts.

In this example, we stick to spin-isotropic channels without any spin-momentum locking, however, experiments with heavy metals that exhibit Giant Spin Hall Effect\cite{niimi2012giant} have been successfully modeled with spin circuits \cite{camsari2015modular,hong2016spin}.

The basic idea of the coupled NLSVs in FIG.~\ref{fi:fig4}a is to engineer a system of LBMs that interact via pure spin currents. To achieve this, a charge current passes through each magnet with a nearby ground. Then, spin-currents polarized in the instantaneous direction of the magnetization of the LBMs are sent toward neighboring LBMs. The key point is to design a system that takes samples from the classical Heisenberg model:
\begin{equation}
   E = - \frac{1}{2} \sum_{i,j} J_{ij}\,(\hat{m}_i\cdot\hat{m}_j)
   \label{eq:Heisenberg model}
\end{equation}
where $J_{ij}$ are the interaction terms and $m_i$ are 3D-magnetization vectors. Finding low-energy (or equilibrium states of the Heisenberg Hamiltonian for probabilities $p_i$ ($\propto \exp(-E/kT)$) is computationally challenging. As such, engineering a system of LBMs to sample from the classical Heisenberg model can be useful for optimization and/or sampling problems. The read-out mechanisms or practical applications of this system is beyond the scope of our discussion here and can be found in Ref.~\cite{bunaiyan2023heisenberg}.

From a modeling perspective, the system shown in FIG.~\ref{fi:fig4}a is quite challenging: one needs to model the NLSV transport where charge and spin-currents are modeled properly. Moreover, a self-consistent solution of stochastic LLG equations with transport modules is needed. In the presence of incoming spin-currents that are transport-dependent, the sLLG equations provide magnetization vectors that control the interface conductances.  The spin-circuit approach allows a seamless implementation of this highly complicated physical system. As shown in FIG.~\ref{fi:fig4}, the magnitude and the sign of the injected charge current control the degree of correlation between two magnets (1 and 2). When the injected charge currents are negative, the coupling between the three LBMs exhibits antiferromagnetic coupling as shown in FIG.~\ref{fi:fig4}c, as would be expected from the engineered interactions obtained from a Heisenberg model with negative couplings.

\begin{figure*}[t!]
    \centering
    \includegraphics[width=0.9\textwidth]{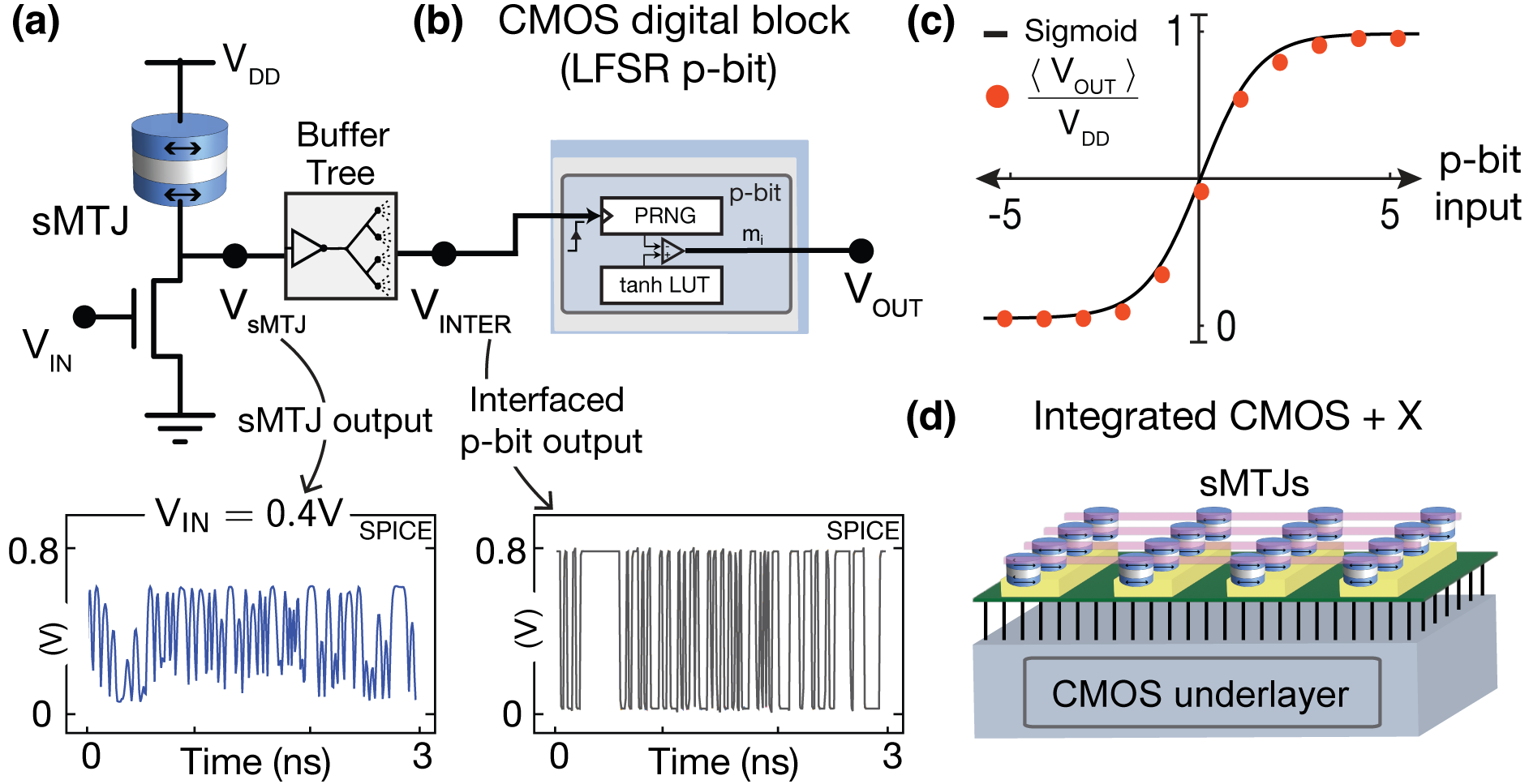}
    \caption{\textbf{CMOS + X (stochastic MTJ) platforms}  (a) An sMTJ-based binary stochastic neuron (p-bit) is interfaced with a digital CMOS-based circuit to trigger a digital p-bit emulator. The bottom panel shows SPICE results for the analog fluctuations at the drain ($\sf V_{sMTJ}$) of the NMOS transistor. 
     (b) Rail-to-rail stochastic fluctuations obtained after a buffer tree ($\sf V_{INTER}$) is inserted between the single sMTJ-based p-bit and the large digital CMOS block.  CMOS block contains a low-quality and inexpensive pseudo-random number generator (PRNG) along with a look-up table to obtain tunability. This hybrid setup with the sMTJ circuit increases the quality of randomness that can be obtained from the digital p-bit block alone (see Ref.~\cite{singh2024cmos})  (c) Tunability of the heterogeneous structure as a probabilistic bit is shown with time-averaged $\sf V_{OUT}$ over 1000 ns in SPICE. (d) In the future, millions of sMTJs can provide nearly-free true randomness to CMOS underlayers for various probabilistic computing applications. 
    }
    \label{fi:fig6}\vspace{-15pt}
\end{figure*}

\section{Functional spin-circuits with transistors}
\label{sec:smtj}
So far, the spin-circuit examples we considered have been based on spintronic building blocks with increasing sophistication, albeit without the use of any transistors. An emerging trend in the field in the beyond Moore era of electronics is the notion of \textit{domain-specific} computation where conventional complementary metal-oxide semiconductor (CMOS) transistors are augmented with emerging technologies ($ \sf X$) to create CMOS+$\sf X$ systems, where $ \sf X$ can stand for anything from spintronics, photonics, memristors, superconducting circuits, and others. In this section, we show how a probabilistic bit with stochastic MTJs combined with CMOS components \cite{borders2019integer, singh2024cmos} can be modeled and characterized within our spin 
circuit formalism. 

FIG.~\ref{fi:fig5} shows how a new type of stochastic magnetic tunnel junction can readily be modeled and analyzed using the spin-circuit approach, in conjunction with state-of-the-art transistor models forming a p-bit building block. Typically, magnetic tunnel junctions employ a fixed layer such that the resistance of the junction correlates with the magnetization of a free layer. With the emergence of probabilistic computing \cite{chowdhury2023full} and the need for fast, energy-efficient, and scalable random number generators, a recent approach has been to design sMTJs with no fixed layers \cite{camsari2021double,sun2023PRBeasyplanedominant,kemal2024double}. 

Typically, physics-based device models cannot easily be interfaced with transistor models, whereas the SPICE formulation of spin-circuits allows seamless integration with CMOS. FIG.~\ref{fi:fig5} shows such a combination where a spin-valve made out of two LBMs is connected to an n-MOS transistor. Here, we use  FinFET models from the open-source predictive technology models (PTM){\cite{zhao2007predictive}}, in principle, however, any other FET model could be combined with spin-circuits. When we combine spin-circuits that carry 4-component currents and voltages with ordinary circuits that only carry charge currents, we only attach the charge current terminals to each other, since any other spin information can be ignored in extended charge circuits. The double-free layer sMTJ exhibits an interesting voltage-independent resistance profile \cite{otakun2024}. This behavior is reproduced by the spin-circuit model where the two symmetric layers receive spin-currents with opposing signs (FIG.~\ref{fi:fig5}b), leading to nearly uniform fluctuations (FIG.~\ref{fi:fig5}c) and weak voltage-bias dependence (FIG.~\ref{fi:fig5}d). Voltage bias independence is shown to be favorable in p-bit circuitry to obtain a clear sigmoidal response in the face of device-to-device variations. The slight asymmetry favoring an anti-parallel configuration stems from the dipolar coupling between the easy-plane magnets which is included in the spin-circuit simulation (following the methodology in Ref.~\cite{camsari2021double}). Later work suggests that building sMTJs out ofsynthetic antiferromagnet (SAF)-based free layers can remove this zero-field dipolar coupling entirely\cite{kemal2024double}. 

Finally, FIG.~\ref{fi:fig5}e-f shows the full input-output characteristics of what has been called a probabilistic or p-bit \cite{camsari2017implementing} that is obtained from our full model. These simulations demonstrate the tunability of randomness at different bias voltages. An important point to stress is that the tunability does not arise from spin-transfer-torque effects modulating the free layer magnetizations, but rather from the changing transistor conductance by an analog input voltage.

 The combination of magnetization dynamics, dipolar and thermal noise fields, 4-component interface conductances, and transistors in a sound powerful circuit simulator shows the power and flexibility of the spin-circuit approach. We believe the approach eases the prediction and device-circuit level evaluation  of new types of spintronic devices. In the case of double-free-layer sMTJs (first with double-free layers\cite{camsari2021double} and then with double-free SAF layers\cite{kemal2024double}), spin-circuit theory predicted the  key qualitative features of these devices that have later been experimentally demonstrated \cite{sun2023easy,otakun2024}. 



\section{From spin-circuits to systems}
\label{sec:cmosx}

So far, the spin-circuit examples we illustrated are all at the device or the circuit level. The combination of spin-circuits with transistors unlocks a much larger space of possibilities including the realization of energy and area efficient p-bit networks. As a final example, we describe a hybrid system where a true random number generator (TRNG) augments a low-quality pseudo-random number generator (PRNG) (even though our modeling in SPICE will use PRNGs for the sMTJ part of this circuit, the RNG quality used for this purpose will be much higher than that of an LFSR without noticeable differences between an actual MTJ, see Ref.~\cite{singh2024cmos} for details). FIG.~\ref{fi:fig6} shows the p-bit circuit from FIG.~\ref{fi:fig5} triggering  a digital p-bit (FIG.~\ref{fi:fig6}a-b) to generate a tunably random behavior (whose average is shown in FIG.~\ref{fi:fig6}c). 

The motivation is to increase the quality of randomness that is extracted from inexpensive linear feedback shift register (LFSR)-based PRNG by clocking the PRNG with random arrivals of sMTJ fluctuations. This setup was realized using physical sMTJs in a recent experiment that established the concept \cite{singh2024cmos}.  Considering the expensive nature of PRNGs, augmenting them with the true randomness of millions of sMTJs in integrated CMOS + $\sf X$ systems (FIG.~\ref{fi:fig6}d) seems desirable. 

The CMOS block consists of a PRNG and a lookup table (LUT) for the hyperbolic tangent function consisting of thousands of transistors. The CMOS design is synthesized from the open-source ASAP7 Predictive PDK~\cite{CLARK2016105} and the details of how this synthesis is performed can be found in Ref.~\cite{singh2024cmos}.

The bottom panels of FIG.~\ref{fi:fig6} show simulations obtained from this hybrid circuit where a single sMTJ-based circuit drives thousands of transistors. An important detail, immediately captured by the spin-circuit approach is that of loading. Without a ``buffer tree'' where several stages of inverters distribute the capacitive load of the CMOS p-bit that is seen by the single sMTJ-based circuit, the clocking does not work. These nontrivial loading effects at the interfaces of physics-based and digital systems are naturally captured by the powerful spin-circuit approach that is otherwise easy to miss. In addition to loading, key circuit and system metrics such as energy-delay can be reliably calculated for many types of exploratory systems.

To verify the functionality of the synthesized p-bit, we probe the drain voltage in FIG.~\ref{fi:fig6}a to observe the sMTJ random telegraph noise when biased at $V_{50/50}$, followed by the rail-to-rail output of the buffer tree in FIG.~\ref{fi:fig6}b. In FIG.~\ref{fi:fig6}c we plot the probability of the p-bit output being 1 against the decimal equivalent of p-bit input, matching the expected sigmoidal behavior. This system represents an energy-efficient and scalable p-bit model, which has demonstrated significant potential in offering scalable solutions to complex previously intractable problems\cite{singh2024cmos, chowdhury2023full}.

\section{Conclusion}
\label{sec:conc}
We have described the spin-circuit approach connecting the microscopic physics of spins and magnets all the way up to circuits and systems. We believe that such a physics-based, modular, CMOS-compatible modeling approach will be critical in evaluating and exploring new hardware systems in the beyond-Moore era of electronics. Despite the wide focus of this paper, there are many other spintronic phenomena that have been modeled by spin-circuits and we did not get into these, e.g. full compute life-cycle modeling of skyrmions and domain walls \cite{sakib2022skyrmion}. Beyond spins, a similar circuit framework for new and emerging phenomena can be constructed. Extensions may include pseudospins, valley currents\cite{hung2019direct}, superconductivity, photonics, and qubit systems \cite{ganguly2023qubitmodeling}. A diverse set of such phenomena can all be analyzed within the context of powerful, industry-standard transistor-compatible circuit simulators, while being rigorously connected to the underlying physics, beyond empirical compact models. In the new era of electronics, such extended spin-circuits could enable a rapid and robust evaluation of emerging CMOS+$\sf X$ systems.  \vspace{-10pt}

\section*{Code and Data Availability}
All the codes used in this study are publicly available in the Github repository [https://github.com/OPUSLab/Spin-Circuit-Designs].

\begin{acknowledgments}
We acknowledge the U.S. National Science Foundation (NSF) support through CCF 2106260, through a CAREER award, SAMSUNG Global Research Outreach (GRO) grant. We also acknowledge support ONR-MURI grant N000142312708, OptNet: Optimization with p-Bit Networks. Use was made of computational facilities purchased with funds from the National Science Foundation (CNS-1725797) and administered by the Center for Scientific Computing (CSC). The CSC is supported by the California NanoSystems Institute and the Materials Research Science and Engineering Center (MRSEC; NSF DMR 2308708) at UC Santa Barbara.

\end{acknowledgments}

\section*{Author Contributions}
KYC and SD wrote the initial draft of the manuscript. KS, SB, NSS have designed the spin-circuits, obtained results, drew the figures and prepared open-source models. All authors have provided feedback and helped improve the manuscript.  

\section*{Competing Interests}
 All authors declare no financial or non-financial competing interests.

\appendix

\section{Modeling stochastic ODEs WITH HSPICE}

Given the circuit description of transport, solving the stochastic magnetization dynamics in standard circuit simulators for transient analysis allows a consistent self-consistent approach. Unfortunately, the LLG equation is nonlinear and cannot be represented as an elegant RLC circuit describing damped oscillations. We resort to the powerful capabilities of HSPICE for handling nonlinear ODEs numerically.  

\begin{figure}[h!]
    \centering
    \includegraphics[width=0.45\textwidth]{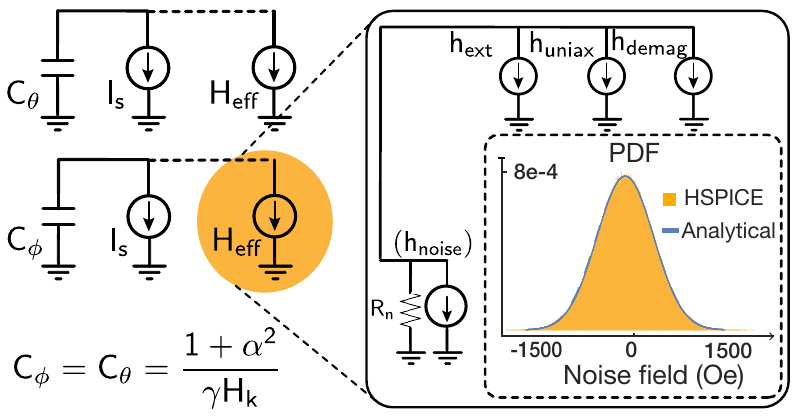}
    \caption{\textbf{Stochastic LLG module and SPICE  circuit description} Capacitor-current source circuits are used to solve for the magnetization components in spherical coordinates $(\theta(t),\phi(t))$. Inset shows the probability density function of the noise field that matches the analytically (Eq.\ref{eq:llgnoise}) with the noise source in HSPICE with minor modifications (see the text).} 
    \label{fi:figAppC1}
\end{figure} 

We implement the sLLG equation (Eq.~\ref{eq:sLLG} )as a capacitor-current source system in 2 dimensions (for $\phi$ and $\theta$): 
\begin{equation}
\underbrace{\left(\frac{1+\alpha^2}{\gamma H_k}\right)}_C \underbrace{\frac{d \hat{m}}{d t}}_{\frac{d V}{d t}}=\underbrace{f\left(\hat{m}, \vec{h}_{e f f}, \vec{I}_s\right)}_{I\left(\hat{m}, \vec{h}_{e f f}, \vec{I}_s\right)}
\end{equation}
where the right hand side of the equations are defined as dependent sources current sources that capture the non-linearities, through their explicit dependence on $\theta$ and $\phi$ and other magnetization parameters. The left hand side is a capacitor whose voltage represents the instantaneous value of $\theta$ and $\phi$. The choice of the spherical coordinates is appropriate since this naturally conserves the magnetization amplitude ($|m|=1$) at all times, however magnetic fields through anisotropies and spin-currents  naturally enter in Cartesian coordinates that are converted to spherical coordinates similar to  the prescriptions discussed in References~\cite{sun2000spin,panagopoulos2013physics,torunbalci2018modular}. 

FIG.~\ref{fi:figAppC1} illustrates the circuit used to obtain $\hat{m}$ where different dependent current sources are organized to represent different terms in the effective magnetic field including uniaxial, demagnetization anisotropies and external magnetic fields. More terms to include new physics such as voltage control of magnetic anisotropy or magnetostriction can easily be added to this module. Special attention needs to be given to the noise terms since the way these terms enter the equation are fundamentally different. Specifically, we rely on the .trannoise function of HSPICE where the basic model is a resistor with a random current source to describe thermal noise. Due to internal, closed-source definitions of HSPICE (using version T-2022.06), we observe that the noise field, representing the magnitude of the autocorrelation of the noise field measured in units of $A^2/Hz$, needs to be defined as ${4 \alpha k_B T}/{M_s \text{Vol.} \, |\gamma| \mu_0}$ which has an extra factor of 2. We carefully checked that when defined with this extra factor, the samples we get correspond to the \textit{correct} probability density function (see inset in FIG.~\ref{fi:figAppC1}). As discussed in the main text, the sLLG module we use has been rigorously benchmarked against a time-dependent Fokker Planck Equation description of magnetization dynamics and is correctly implemented by HSPICE.

\begin{figure}[t!]
    \centering
    \includegraphics[width=0.45\textwidth]{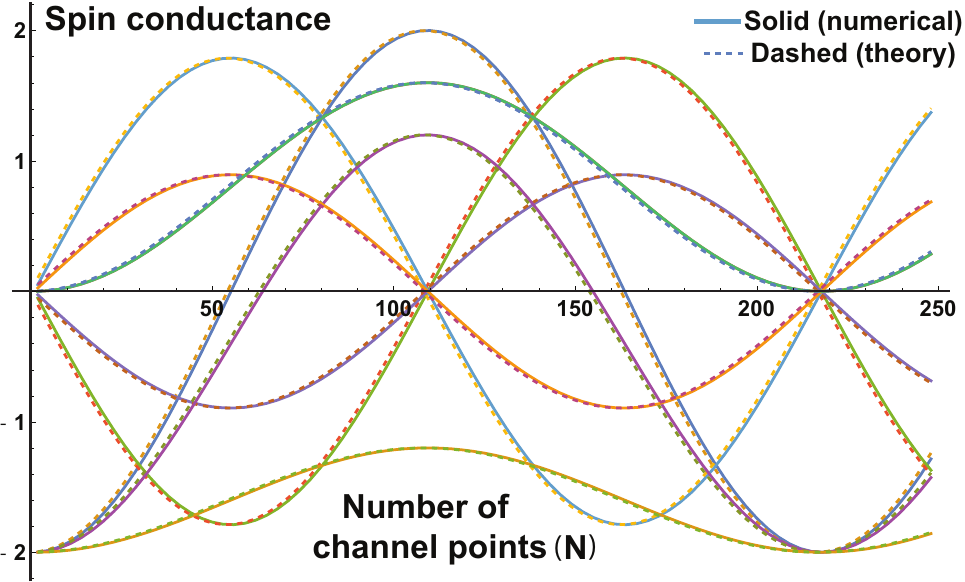}
    \caption{\textbf{Spin Conductances for RSO + DSO channel in 1D:} we compare the theoretical spin submatrix ($zz,zx,zy,xz,xx,yz,yx,yy$) components of Eq.~\ref{eq:gsoc} against a numerical NEGF calculation, showing excellent agreement. The parameters are $\alpha=5\times 10^{-11}$ eV-m, $\beta=2.5 \times 10^{-11}$ eV-m, lattice spacing $a=0.1$ nm, the injection energy is equal to $t$, where $t=\hbar^2/(2 m^* a^2 q)$ where $q$ is the electron charge and $m^*$ is the effective mass with $0.2 m_e$ where $m_e$ is the electron mass. The rotation angle is $\theta=\sqrt{\alpha^2+\beta^2} 2 m^* (N a) q/\hbar^2 $ where $N a$ is the channel length. } 
    \label{fi:soc}
\end{figure} 

\section{From Non-Equilibrium Green's Function Formalism to 4-component spin conductors}

In this section, we will show the derivation of conductance matrix corresponding to a channel with Rashba spin-orbit (RSO) and Dresselhaus spin-orbit (DSO) coupling (Eq.~\ref{eq:gsoc}). Our starting point is Hamiltonian:
\begin{equation}
H = H_0 + \alpha (\sigma_x k_y - \sigma_y k_x) + \beta (\sigma_x k_x + \sigma_y k_y)
\label{eq:hamiltonian}
\end{equation}
Consider a 1D channel with $N=3$ points.  In the tight-binding approximation\cite{datta2018lessons}, we set the hopping parameter and lattice constant a to 1 to ease our analytical calculation: $t=\hbar^2/(2m^* a^2 q)\equiv 1$, $a=1$. The Hamiltonian then reads:
\begin{equation}
\footnotesize
\begin{bmatrix}
2 & 0 & -1 & \frac{1}{2}(-\alpha + i\beta) & 0 & 0 \\
0 & 2 & \frac{1}{2}(\alpha + i\beta) & -1 & 0 & 0 \\
-1 & \frac{1}{2}(\alpha - i\beta) & 2 & 0 & -1 & \frac{1}{2}(-\alpha + i\beta) \\
\frac{1}{2}(-\alpha - i\beta) & -1 & 0 & 2 & \frac{1}{2}(\alpha + i\beta) & -1 \\
0 & 0 & -1 & \frac{1}{2}(\alpha - i\beta) & 2 & 0 \\
0 & 0 & \frac{1}{2}(-\alpha - i\beta) & -1 & 0 & 2 \\
\vspace{-3pt}\end{bmatrix}
\nonumber
\end{equation}
The NM leads on the left and right are described the following self-energy matrices: 
\[
\Sigma = 
\begin{cases}
\Sigma_L(i,j) = -te^{ika} \left[(\delta_{i,1} + \delta_{i,2})\delta_{i,j}\right], \\ 
\Sigma_R(i,j) = -te^{ika} \left[(\delta_{i,5} + \delta_{i,6})\delta_{i,j}\right] 
\end{cases}
\]
here $\delta_{i,j}$ is the Kronecker delta, $k$ is the wavevector at a given energy and $a$ is the lattice spacing. We can then compute the retarded Green's function, $G^R$, via $G^R = [E I - H - \Sigma_L - \Sigma_R]^{-1}$ at a given energy $E=2 t [1- \cos (k a)]$ that approximates a parabolic dispersion when $(k a) \ll 1$. The $G^R$ produces a dense 6$\times$6 matrix that we do not show here. Once $G^R$ is computed, we set $E=t$ and compute the conductance matrices using Eq.~\ref{eq:negf}. Below, we only report $G_{LR}=G_{12}$ for $\alpha/t,\beta/t \ll 1$, which is the  appropriate limit for the perturbative SOC Hamiltonian. We obtain $G_{12}/G_0$: 
\begin{equation}=-\begin{bmatrix}
\begin{array}{cccc}
 1 & 0 & 0 & 0 \\
 0 & -2 \alpha ^2-2 \beta ^2+1 & 2 \alpha  & 2 \beta  \\
 0 & -2\alpha  & 1-2 \alpha ^2 & -2 \alpha  \beta  \\
 0 & -2 \beta  & -2 \alpha  \beta  & 1-2 \beta ^2 \\
\end{array}
\end{bmatrix}\label{eq:small}
\end{equation}
where $G_0$=$2 q^2/h$ as defined earlier. It can be readily checked that assuming $\theta=\sqrt{\alpha^2+\beta^2}(L)$ where $L = (N-1)$, $N-1$ being the channel length and using $\gamma=\rm tan^{-1}(\beta/\alpha)$, the small $\alpha,\beta$ expansion of the rotation matrix shown in the main text reproduces Eq.~\ref{eq:small}, proving the relation. The $(N-1)$ factor for the channel length ensures there is no rotation for a channel length of 1, which would consist of a single $2\times 2$ spin site without any spin-orbit interaction. FIG.~\ref{fi:soc} shows a numerical comparison of a 1D NEGF showing excellent agreement with spin conductance components of Eq.~\ref{eq:gsoc}.



\bibliographystyle{unsrtnat}



\end{document}